# Alignment and Characterisation of Remote-Refocusing Systems

**WENZHI HONG,[1] HUGH SPARKS,[1] AND CHRIS DUNSBY[1,*]**

[1]*Photonics Group, Physics Department, Imperial College London, London, UK*
*\*christopher.dunsby@imperial.ac.uk*

**Abstract:** The technique of remote refocusing is used in optical microscopy to provide rapid axial scanning without mechanically perturbing the sample and in techniques such as oblique plane microscopy that build on remote refocusing to image a tilted plane within the sample. The magnification between the pupils of the primary (O1) and secondary (O2) microscope objectives of the remote-refocusing system has been shown previously by Mohanan and Corbett [J Microsc, **288**(2):95-105 (2022)] to be crucial in obtaining the broadest possible remote-refocusing range. In this work, we performed an initial alignment of a remote-refocusing system and then studied the effect of axial misalignments of O1 and O2, axial misalignment of the primary tube lens (TL1) relative to the secondary tube lens (TL2), lateral misalignments of TL2 and changes in the focal length of TL2. For each instance of the setup, we measured the mean point spread function $FWHM_{xy}$ of 100 nm fluorescent beads, the normalised bead integrated fluorescence signal, and calculated the axial and lateral distortion of the system: all of these quantities were mapped over the remote-refocusing range and as a function of lateral image position. This allowed us to estimate the volume over which diffraction-limited performance is achieved and how this changes with the alignment of the system.

## 1. Introduction

The technique of remote refocusing introduced by Botcherby et al. [1] consists of three microscopes in series. The first two microscopes consist of objectives O1 and O2 and tube lenses TL1 and TL2 respectively. These microscopes are arranged in a back-to-back configuration, with the overall magnification set to match the ratio of the refractive indices in the sample and intermediate spaces. This produces an intermediate image with equal lateral and axial magnification. The third microscope then provides a magnified image of a user-defined plane within the intermediate image, which can be refocused away from the focal plane of the first microscope objective. This allows the focal plane of a high numerical aperture objective microscope system to be scanned without mechanically disturbing the sample. Remote refocusing has found numerous applications in high-speed 2D and 3D multiphoton microscopy [2, 3]. It has also been applied in various other imaging techniques, such as spinning disk-remote focusing (SD-RF) microscopy [4], remote-refocusing light-sheet microscopy [5] and oblique plane microscopy (OPM) where remote refocusing is used to generate an image of a tilted plane in the sample [6]. The effectiveness of remote refocusing has been demonstrated in diverse biological systems, including imaging of neural circuits [7], embryonic development [3], and dynamic imaging in isolated cardiomyocytes [8].

Prior research into the performance of remote-refocusing systems has primarily investigated the impact of overall magnification, spherical aberration and objective collar-corrected residual aberrations [9, 10]. However, no analysis of the performance of remote-refocusing systems both on and away from the optical axis and in the presence of misalignments of different optical elements has yet been conducted. In order to better understand how different alignment parameters affect the performance of remote-refocusing systems, a folded remote-refocus system (O1 = 60x/1.2NA water and O2 = 50x/0.95NA air) employing a star test mask (STM) and 100 nm diameter fluorescent beads as test objects was developed. We demonstrate the effect of different misalignments of O1, O2 and TL2 by analysing the system distortion using the STM and the system resolution through the full width half maximum (FWHM) measurements of 100 nm beads within a refocusing range from -100 μm to 100 μm. These misalignments were chosen as they represent the degrees of freedom in the optical systems used routinely in our laboratory. Our findings highlight the critical role of proper alignment in the performance of remote-refocusing systems. The results illustrate the 3D region over which diffraction-limited remote refocusing is achieved both on and away from the optical axis for a specific remote-refocusing system implementation.

## 2. Methods

### 2.1 Magnification and alignment requirements of remote-refocusing systems

In remote-refocusing systems, to achieve diffraction-limited imaging of a sample away from the design focal plane of a primary objective, the output from the primary microscope (O1 and TL1) is passed into a secondary microscope (O2 and TL2), as shown in Fig. 1 (a) and (b). The magnification of the secondary microscope is chosen so that the overall lateral magnification from the sample to the intermediate image formed at the output of O2 is equal to the ratio of the refractive indices of the immersion media used in the sample ($n$) and intermediate image ($n'$) spaces [1].

For instance, in a water immersion O1 imaging system, the remote-refocusing system requires the overall lateral magnification ($M_{lateral}$) to be $n/n' = 1.33$. Additionally, the axial magnification ($M_{axial}$) in this scenario is also equal to 1.33, as determined by the formula for lateral and axial magnifications ($M_{axial} = M_{lateral}^2 \, n'/n$). In addition to the requirement of the overall lateral magnification, remote-refocusing systems should also meet the conjugate pupil condition that the pupil image of O1 is accurately projected the pupil plane of O2 to eliminate misalignment effects [9]. Therefore, an investigation of the misalignment effects induced by O1, O2, and TL2 is performed here to help understand how to obtain the best performance in remote-refocusing systems.

### 2.2 Optical setup

The configurations of the transmitted light and epi-fluorescence imaging modes of the test-rig system are shown in Fig. 1(a) and Fig 1(b) respectively. The magnifications of the tested primary and secondary microscopes are chosen as the same as the remote-refocusing system in [11], as well as System I in [12].

In the transmitted-light mode, light from a 530 nm LED light source (M530F2, Thorlabs) is directed through an aspheric lens (AL, 350230-B, Thorlabs) with a focal length of 4.5 mm to create an illumination spot on a Lambertian diffuser (LD, 50 DO 50, Comar Optics). The size of the illumination spot on the diffuser was chosen to try and ensure that the illumination NA exceeds that of O1. The distance from the first LD surface to the front surface of STM is 3.5 mm, therefore we required the diameter of the illumination spot to be larger than 14.5 mm, and hence a spot size of ~20 mm was used. The alignment object, a star test mask (STM, JD Photo Data), consists of a hexagonal array of 1 μm diameter pinholes in a chrome coating with a spacing of 20 μm. A 0.17 mm (#1.5) thick precision coverslip (630-2186, Marienfield) is placed in contact with the STM, and a drop of water is placed between them. The correction collar for O1 is then set to near the 0.17 mm coverslip position so as to provide symmetrical defocus around the focus position. The LD and STM are placed in contact and mounted on an XYZ translation stage. The XY motion is provided by a manually controlled stage (XY Stage, XYT1/M, Thorlabs) that is mounted on a motorised stage (Stage 1, M-UMR8.25 and TRA25CC, Newport) that provides Z motion to achieve sample refocusing with a minimum actuator step size of 0.2 μm. The front surface of Stage 1 was aligned to be normal to the optical axis of the system. This was achieved by placing a plane mirror flush on the surface of Stage 1. O1 was removed and an alignment laser passed through a pinhole centred in the optical rail system. Stage 1 was then adjusted so that the beam is retroreflected by the mirror back through the pinhole. The coverslip was held in place by placing a drop (50 µl) of water between the STM and the coverslip. Surface tension held the coverslip in place and the drop volume so that the wetted area was smaller than the area of the coverslip. This ensured that the coverslip was as close and parallel as possible to the STM. Following imaging through the objective (O1, 60×, 1.2NA water immersion, MRD07602, Nikon), the beam proceeds through a 4-f system incorporating TL1 ($f$ = 200 mm, ITL200, Thorlabs) and TL2 (composed of D1 ($f$ = 300 mm, 322305000, Linos) and D2 ($f$ = 300 mm, AC508-300-A, Thorlabs)). The separation of D1 and D2 are adjusted as described in the alignment procedure below to achieve a focal length of ~162.4 mm. The beam then propagates towards the secondary objective (O2, 50×, 0.95NA air, MPLAPON, Olympus) via a polarising beamsplitter (PBS, 49002, Edmund Optics) and a quarter-wave plate (QWP, AQWP10M-580, Thorlabs) with a 3°53' tilt (using a pair of shim plates, SM1W353, Thorlabs) to avoid unwanted back reflections. A refocusing mirror (M2, PF10-03-G01, Thorlabs) is placed near the focal plane of O2 that is also driven by a motorised stage (Stage 2, M-UMR8.25, TRA25CC, Newport). The reflected beam is finally collected via O2 and TL3 ($f$ = 200 mm, TTL200A, Thorlabs) to a sCMOS camera (C2, ORCA-Fusion, Hamamatsu). M1 in Fig.1 (a) is a detachable mirror (PF10-03-G01, Thorlabs) and is used to direct light from the sample to TL4 (TTL100A, Thorlabs) and a CMOS camera (C1, MQ013CG-E2, Ximea) that is used during positioning of the sample into the focal plane of O1.

For the epi-fluorescence imaging mode (Fig. 1 (b)), a sample of 100 nm diameter fluorescent beads (T7279, TetraSpeck) is used as the imaging object (see Supplementary Material section 1 for sample preparation). Epi-fluorescence illumination is achieved by imaging LED2 (M470L2, Thorlabs) to the pupil plane of O1 via a lens pair (L1, LA1401-A and L2 AC508-200-A-ML, both Thorlabs) and a long-pass 495 nm dichroic beamsplitter (DB, 25FF495-Di03, Semrock). An excitation filter (F1, FF01-466/40-25, Semrock) and an emission filter (F2, FF03-525/50-25, Semrock) are placed respectively in front of LED2 and TL3 to prevent excitation light from LED2 reaching camera C2.

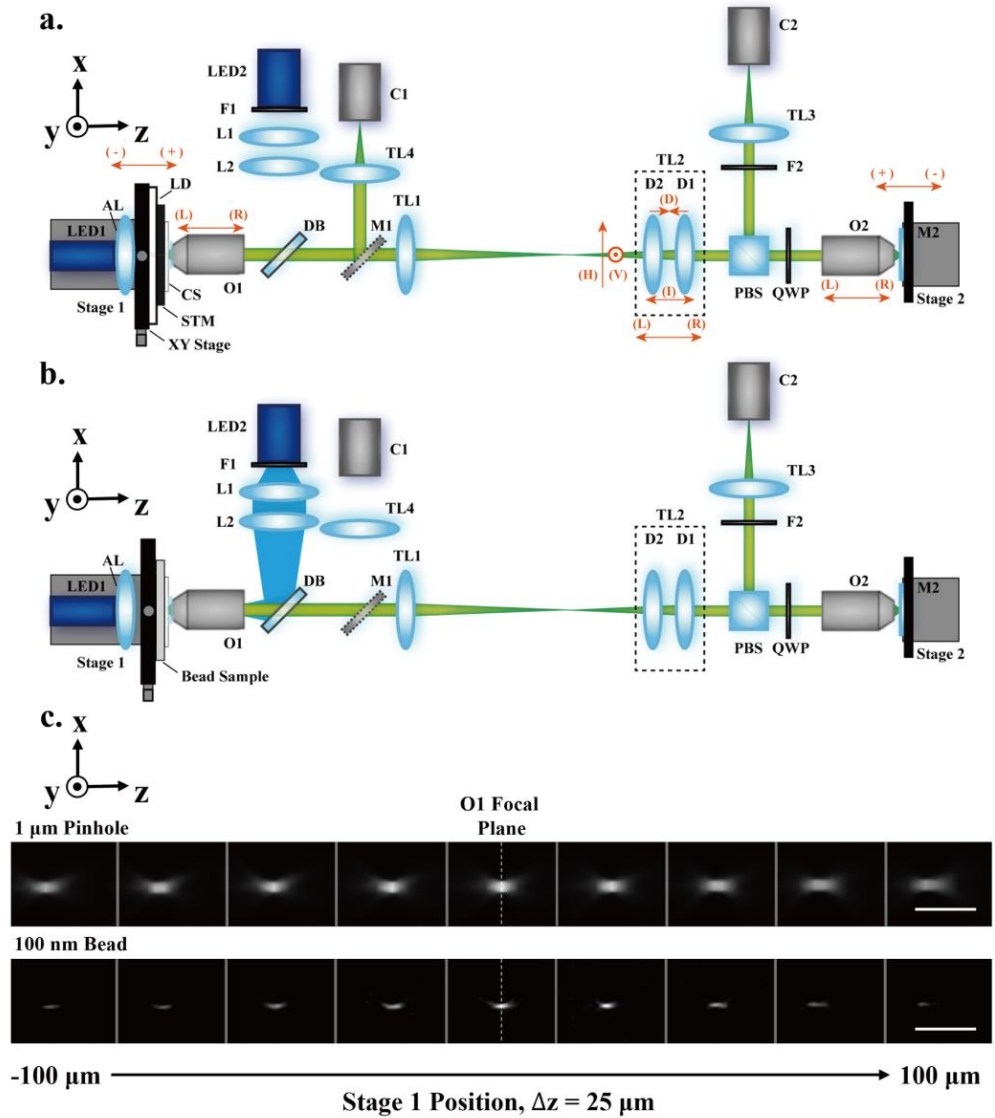

Fig. 1. Configurations of the transmitted light and epi-fluorescence imaging modes of the test rig system. (a) Schematic of the transmitted-light mode. (b) Schematic of the epi-fluorescence imaging mode. (c) Top, exemplar zx-view of sub-regions of interest (sub-ROIs) of the same pinhole at different Stage 1 positions across the whole refocusing range acquired using the transmitted-light mode. Bottom, same for a single fluorescent bead acquired using the epi-fluorescence mode. Vertical grey lines separate the images taken at different Stage 1 positions. The white dashed vertical line indicates the position of the focal plane of O1. Scale bar, 5 μm. AL, aspheric lens; O, objective; TL, tube lens; M, mirror; L, lens; C, CMOS; D, doublet; F, filter; LD, Lambertian diffuser; STM, star test mask; CS, coverslip; DB, dichroic beamsplitter; PBS, polarising beamsplitter; and QWP, quarter-wave plate. Red arrows in (a) correspond to the stage directions and component misalignment directions.

### 2.3 Initial system alignment

Initial alignment of the remote-refocusing setup was performed using the protocol used in our laboratory:

1. The distance between the back focal plane of O1 and the first principal plane of TL1 was set to the focal length of TL1 using a collimated laser diode (PL202, Thorlabs) and shear plate (SI050, Thorlabs).
2. The distance between TL1 and TL2 was set to equal the sum of their focal lengths using the collimated laser diode and shear plate.
3. The initial axial O2 position relative to TL2 was set using the shear plate and collimated laser diode, with the diode incident on the front surface of O2.
4. Then add TL3 and camera C2, with their separation pre-set at the TL3 focal length by imaging an object at infinity.
5. Align the lateral position of TL2 by using STM or USFA test chart as the sample and moving the object to a defocus position of 100 μm with respect to the O1 focal plane. Mark the position of a chosen feature of the object in the C2 image, move the object to the other side of O1 focal plane at -100 μm and refocus the image on C2 using the remote mirror M2 on Stage 2. Adjust the lateral position of TL2 to overlap the object image with the marked position. Five iterations of this process are generally sufficient to remove all measurable lateral motion of the image under changes in refocus.
6. Use the procedure described in section 2.5 to measure the lateral magnification of the system as a function of refocus and the axial magnification of the system.
7. Use the gradient of the lateral magnification to adjust the axial position of O2. In this work, we chose based on experience to define the initial alignment as one where the absolute gradient was less than $5\times10^{-5}$ μm$^{-1}$. A positive gradient larger than $5\times10^{-5}$ μm$^{-1}$ means that O2 should be moved away from TL2, while a negative gradient smaller than $-5\times10^{-5}$ μm$^{-1}$ means that O2 should be moved towards TL2. Return to Step 5. If no adjustment proceed to Step 8.
8. Check whether the system lateral and axial magnifications agree to within 0.01 (This criterion was chosen for the initial alignment based on experience, and considering that for OPM imaging it is important to obtain equal lateral and axial magnifications.) If they do not agree, then adjust the focal length of TL2 and return to Step 5. Once the variation in lateral magnification as a function of refocus (absolute gradient) < $5\times10^{-5}$ μm$^{-1}$, and lateral and axial magnifications matching to within 0.01 then stop.

Following this procedure, the lateral magnification was 1.3324±0.0005 (gradient = $(2\pm0.8)\times10^{-5}$ μm$^{-1}$) and the axial magnification was 1.334±0.005 (error bars report random error): this was defined as the initial alignment of the system against which misalignments were compared. We note that the process of checking the lateral magnification as a function of remote-refocusing distance was described previously by Yang et al. and Kim et al. [13, 14].

*2.4 Control software and acquisition of system characterisation image data*

Translation of the stages and image acquisition were controlled by custom-written LabVIEW software. For each Stage 1 (sample defocus) position, a stack of images is acquired at different Stage 2 (remote refocus) positions. The resulting image data allows for analysis of the characteristics of the remote-refocusing system. When imaging the STM in transmitted light mode, Stage 1 is incrementally moved from -100 μm to 100 μm with 25 μm intervals. At each Stage 1 position, Stage 2 is repositioned to the nominal refocusing plane and scanned from 3 μm to -3 μm around this position with -0.5 μm steps (117 images in total). Exemplar zx sub-ROIs for one pinhole are shown in the top row of Fig. 1 (c). When imaging the fluorescent beads in epi-fluorescence imaging mode, Stage 2 is scanned over the same range of 3 μm to -3 μm but with the minimum incremental motion of the motorised actuator (-0.2 μm), resulting in 279 images (see bottom row of Fig. 1 (c) for an exemplar of zx sub-ROIs for one bead). The positive and negative directions of stage motion are indicated in Fig. 1 (a).

*2.5 Characterisation of lateral and axial magnification*

The lateral and axial magnifications from the sample to the remote space – i.e. the combined magnification of microscopes 1 and 2 – were determined from images of the star test mask in transmitted light mode with 0.5 μm steps of Stage 2. For each Stage 1 (sample defocus) position, a square ROI with side 60 pixels was defined for each pinhole within the centre 67×67 μm$^2$ of the field of view (FOV) and used to find the maximum pixel value of each pinhole. The image with the highest average maximum pinhole pixel value was then selected as the in-focus image, which corresponds to the Stage 2 position closest to where the STM is in focus on C2.

The central location of each pinhole in each in-focus image was determined by first subtracting a background, which was estimated by morphologically opening the raw image with a disk structuring element (radius of 15 pixel), and then applying a threshold (twice the average of the estimated background) to the image for binarisation [15]. Pinhole regions were then found by finding connected components in the binary image. The position of each pinhole was then determined from the centroid of its image within each connected component. The average of adjacent pinhole separations within the centre of FOV (67×67 μm$^2$) where the lateral distortion is relatively small and stable (see Fig.2 (d)), was calculated to estimate the lateral magnification for each Stage 1 position. This process was repeated for all 9 Stage 1 positions over the range from -100 μm to 100 μm and the lateral magnification was plotted as a function of Stage 1 position, see Fig. S1 (a) for an example measured using the initial alignment of the system. A linear function was then fit to the data to determine the gradient and y-intercept using MATLABs *fitlm* function, which returns standard deviations on the fit parameters. The y-intercept obtained from this plot provided the overall lateral magnification of the whole system (microscopes 1, 2 and 3).

In order to determine the combined magnification of just microscopes 1 and 2, it was then necessary to determine the lateral magnification of microscope 3 alone. This was achieved by removing the STM from the front focal plane of O1, removing mirror M2 and placing the STM in the front focal plane of O2. No coverslip was used as O2 is designed to operate without coverslip. The transillumination optics (LED1, AL1 and LD) were also repositioned to illuminate the STM. Thirteen in-focus images of the STM were acquired with the STM translated laterally by a random amount between images. For each in-focus image, the lateral magnification of microscope 3 was calculated using the method described above, which was found to be 55.55±0.02 (mean ± standard error). The combined lateral magnification of microscopes 1 and 2 was then found by dividing the magnification of microscopes 1,2 and 3 by the magnification of microscope 3. The use of the same physical STM and camera to determine the lateral magnification of the overall system and the lateral magnification of microscope 3 mean that the STM pinhole spacing, and camera pixel dimensions cancel out overall, thus avoiding any potential systematic errors due to uncertainties on these quantities.

To determine the axial magnification, two times the Stage 2 positions from the 9 in-focus images were plotted as a function of Stage 1 position. (The Stage 2 position is doubled to give the optical path length change on reflection.) A linear fit is applied to the data, again using MATLABs *fitlm* function, and the gradient provides the negative of the axial magnification (see Fig. S1 (b) for an example of the initial system).

The random error on the measurement of the lateral magnification, axial magnification, and gradient of lateral magnification with refocus distance was taken to be the average standard deviations from the *fitlm* fitting function over all measurements presented in this work. This yielded an error of 0.0005 for the lateral magnification, 0.005 for the axial magnification and 8×10$^{-6}$ for the gradient of the lateral magnification as a function of sample refocus distance. These error estimates are relevant when comparing between different values reported within this paper.

The relative random error on the measurement of the magnification of microscope 3 leads to a systematic error in the reported lateral magnifications of microscopes 1 and 2 of 0.0005. This error affects all lateral magnification measurements reported in this paper equally, and so is not relevant when comparing between values reported here, but it is relevant when comparing to values measured in different experimental implementations or when comparing the measured lateral magnification to the ideal value calculated from the ratio of refractive indices between sample and intermediate image spaces.

*2.6 Characterisation of system distortion*

As the 1 μm diameter pinholes in the STM are arranged in a hexagonal array, it is possible to estimate the lateral distortion introduced by the refocusing system by calculating the distance between each pinhole in an in-focus image of the STM (mentioned in section 2.5) to the nearest pinhole in a predetermined reference pinhole hexagonal array. The pinhole closest to the FOV centre in the in-focus image at 0 μm sample defocus position was selected as the centre predetermined reference pinhole. The system lateral magnification measured in section 2.5 was then used to generate the predetermined reference pinhole hexagonal array. The 9 in-focus images at different Stage 1 positions were analysed to calculate the lateral distortion (see Fig. S2). These values were then utilised to generate a lateral distortion map as a function of the distance from the FOV centre (see Fig. 2 (c)).

The Stage 2 position of the in-focus image obtained for a Stage 1 position of 0 μm is used to determine the zero position of Stage 2, and this value is then subtracted from all Stage 2 positions.

To measure the axial distortion, 7 images with the highest average maximum pixel value of pinholes within the FOV centre were selected at each Stage 1 (sample defocus) position. Field curvature maps were then obtained by considering each individual pinhole. The axial distortion for a pinhole was determined by the displacement between the nominal Stage 2 position and the position that produced the highest estimated Strehl ratio for that pinhole, see also [16, 17]. To account for the double-pass effect of M2 and the nominal axial magnification of the system, the displacements were converted into axial distortion in sample space by multiplying a factor of $2/1.333 = 1.5$. An axial distortion map for the 9 Stage 1 positions was then generated as a function of the distance to the FOV centre (see Fig. 2 (d)).

The lateral magnification and intermediate image defocus as a function of sample defocus under axial misalignments of O1, O2 and TL2 were simulated in Zemax OpticStudio (version 20.2) by modelling O1, TL1, TL2 and O2 as Zemax paraxial thin lenses.

*2.7 Characterisation of system resolution*

To characterise the system resolution, epi-fluorescence images of the 100 nm fluorescent bead sample were acquired with the finer Stage 2 step size of 0.2 μm. By considering the NA of the system of 1.2 (limited by O1), an emission wavelength of 525 nm, the 100 nm bead diameter, the 6.5 μm pixel size, the system magnification of 74.0× and the Strehl condition, an estimated diffraction-limited lateral resolution with full width at half maximum (FWHM) of 0.26 μm was calculated, see Supplementary Material section 2.1.

For each Stage 1 (sample defocus) position, 31 images were acquired corresponding to an interval of 0.3 μm in sample space (see example data from one bead shown in Fig. 1 (c)). At each sample defocus position, a difference of Gaussians (DoG) filter (sigma values of 1.0 and 1.46 pixels) [18] was applied and followed by a binarisation step and identification of connected components. The threshold was chosen by gradually increasing the threshold until the number of detected components reached a steady value. The 3-dimensional coordinates (x, y, z) of each bead were then found from the location of the pixel with maximum value within each connected component mask. A preliminary analysis was conducted first to find the average peak bead intensity. To reject bead clusters or hot pixels, the formal analysis was then carried out using an intensity filter that only selects beads with peak intensity ranging from two thirds to four thirds of the average peak intensity. Horizontal and vertical line profiles were taken through each bead with a line length of 1.67 μm in sample space, and the lateral FWHM values were calculated directly using linear interpolation of these profiles. We report the mean FWHM of these two values as the lateral FWHM value of the bead image (see Fig. S4).

We also calculated the integrated energy (pixel sum) for each bead. This measurement requires a choice of the region over which the integration is performed. A larger region will collect a greater fraction of the energy, but also be more likely to be biased by signal from neighbouring beads. It is also necessary to choose a region around the bead over which the local background signal level is estimated. We chose to integrate the energy of the bead over a circle that contains 91% of the energy for an ideal (diffraction-limited) PSF, which corresponds to a circle of diameter 0.88 μm (equivalent to 1.64 Airy units). The background signal level was determined by taking the average of the region outside the circle used to measure the bead energy and within a square of size 1.67 μm in sample space.

To visualise the system performance over a 3D region, the mean lateral FWHM ($FWHM_{xy}$) of beads was analysed to produce a 2D histogram (displayed as a heatmap) with contours (see Fig. 2 (a)). The diffraction-limited volume (see Table S3) was then obtained by finding the volume encompassed by the bins of the 2D histogram with mean $FWHM_{xy} \leq 0.26$ μm in the mean lateral FWHM ($FWHM_{xy}$) map (Fig. 2 (a)). A normalised bead integrated energy map (Fig. 2 (b)) was also generated. The volume in sample space where the normalised bead integrated energy was $\geq 0.8$ is presented in Table S4. Both maps were plotted as functions of sample defocus and bead distance to the FOV centre.

### 2.8 Measurement of the refractive index of immersion water

An Abbe 5 refractometer (Bellingham and Stanley) was used to make 5 repeated measurements of the refractive index of the water used as the immersion fluid. This water was obtained from a water purification system (Type 1 water, ZRQSVR5WW, Direct-Q, Milli-Q). A value of 1.3335±0.0001 (mean ± standard error) for a wavelength of 589 nm, 21°C and atmospheric pressure was determined.

## 3. Results

### 3.1 Axial misalignments of O1 and O2

To investigate the impact of O1 and O2 axial misalignments on remote-refocusing systems, the O1 position in the initial setup was shifted by 1 mm towards TL1 (denoted as (R) in Fig. 1(a) and Fig. S5), as well as 1 mm away from TL1 (denoted as (L) in Fig. 1(a) and Fig. S5). O1 was returned to its initial position and then O2 was adjusted from its position in the initial setup by 2 mm towards TL2 (denoted as (L) in Fig. 1(a) and Fig. 2), as well as 2 mm closer to M2 (denoted as (R) in Fig. 1(a) and Fig. 2). The STM and 100 nm bead samples were imaged under these four misalignments to characterise the system distortion and resolution (see Fig. 2 and Fig. S5). Exemplar bead images under misalignment O2 L are shown in Supplementary Fig. S6.

The plots of lateral magnification measured for different O1 axial position are shown in Fig. S5 (m). Moving O1 towards TL1 (R) causes the pupil of O1 to be imaged to the right of the pupil of O2. Therefore, axially misaligning O1 away from its position in the initial setup causes the gradient of the lateral magnification as a function of sample defocus to decrease and become negative when moved to the left (L) and increase when moved to the right (R) as expected, see also Table S2. Axially misaligning O2 has a similar effect to O1 axial misalignment but the direction of the change in gradient with respect to direction of movement of O2 is reversed, see Fig. 2 (m), again as expected.

The intermediate image defocus, which is twice the relative mirror M2 motion between the actual refocusing position and the ideal refocusing position, as a function of sample defocus are shown in Fig. 2 (n) and Fig. S5 (n) and were fitted to second-order polynomials. The magnitude of the quadratic term increases compared to the initial alignment when O2 was adjusted by 2 mm away from TL2 (R) and decreases and becomes negative when O2 was adjusted by 2 mm towards TL2 (L), see Fig. 2 (n). A similar effect occurs in Fig. S5 (n) when O1 was shifted

1 mm closer to TL1 (R), but here the magnitude of the quadratic term for this misalignment is lower than that of the initial system.

The axial misalignment of O2 in the two directions has approximately equal and opposite effects on lateral and axial distortion (compare Fig. 2 (g, k and h, l)) and on average increases system lateral distortion, see Table S3. Average axial distortion is increased for motion to the right (R) and marginally decreased for motion to the left (L). As the axial misalignment of O1 is smaller (1 mm compared to 2 mm for O2), the effect on distortion is more subtle. Motion of O1 to the left (L), see Fig. S5 (k, l), results in a similar effect to the motion of O2 to the right (R). The motion of O1 to the right (R) decreases the average lateral and axial distortion (see Table S3). It seems likely that this misalignment introduces a negative axial distortion that partially compensates for the axial distortion in the initial alignment, see Fig. S5 (n), that is expected theoretically [19]. However, this came at the cost of a reduction in diffraction-limited image volume characterisation, as the axial misalignments of O1 and O2 reduce the diffraction-limited volume by more than 65% (see Table S3), which can also be seen by the smaller region over which a mean $FWHM_{xy} \leq 0.26$ μm is achieved, see Fig. 2 (a, e, i) and Fig. S5 (a, e, i). The normalised bead integrated energy as a function of sample defocus also becomes slightly narrower – compare Fig. 2 (b) with Fig. 2 (f, j) and Fig. S5 (b) with Fig. S5 (j) – but not in the case O1 R, see Fig. S5 (f). See also the volume over which the normalised bead integrated energy $\geq 0.8$ shown in Table S4.

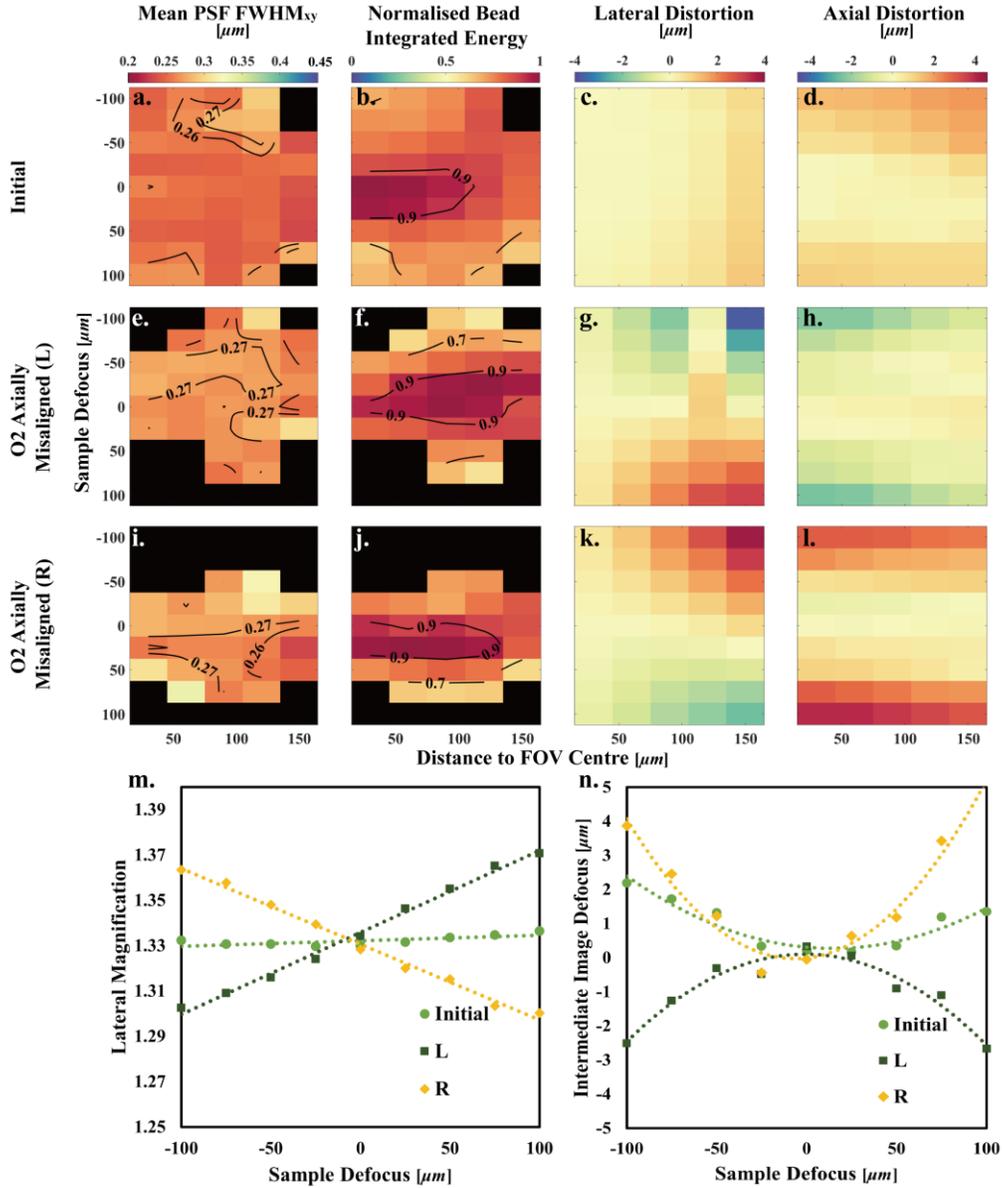

Fig. 2. Characterisation of O2 axial misalignment. (a-d) Initial system. (e-h) O2 axially misaligned by 2 mm towards TL2. (i-l) O2 axially misaligned by 2 mm towards M2. Black blocks indicate no bead recognised in those regions. Fig. S6 shows exemplar bead images for a Distance to FOV Centre of 40 μm and for sample defocuses of 0 and 100 μm for the case of O2 L. (m) Lateral magnification as a function of sample defocus. The dashed lines are the linear fits to the measured data points. (n) Intermediate image defocus as a function of sample defocus. The dashed curves are the second-order polynomial fits to the measured data points.

### 3.2 Axial and lateral misalignments of TL2

As stated in section 2.3, the axial distance between TL1 and TL2 was aligned via a shear plate so they are separated by the sum of their focal lengths. Shear plate alignment error was estimated to be 1.54 mm via the standard deviation of ten measurements of TL1 with respect to TL2. Therefore, the axial misalignments of TL2 were conducted by positioning TL2 2 mm

(larger than the shear plate alignment error) to the left and right from the initial position. In both TL2 axial misalignments, O2 was also realigned to reduce the variation in lateral magnification with defocus to less than $5\times10^{-5}$ $\mu m^{-1}$, as per the initial alignment procedure (see Fig. 3 (m) and Table S2). The mean $FWHM_{xy}$ maps in Fig. 3 (e) for TL2 (L) and Fig. 3 (i) for TL2 (R) are worse than the initial system (smaller diffraction-limited volumes, decreased by 75% and 19%, see Table S3), but the volume over which the normalised bead integrated energy $\geq 0.8$ are slightly higher (increased by 32% and 3%, see Fig. 3 (f, j) and Table S4). Under TL2 axial misalignments in both directions, the mean absolute lateral distortion decreases from the initial alignment, while the mean absolute axial distortion decreases for one direction (L) and increases for the other (R) see Fig. 3 (c-d, g-h and k-l) and Table S3.

Lateral misalignments of TL2 were achieved by adjusting the cage system translating lens mount (CXY2, Thorlabs) horizontally and vertically by 0.5 turns of the relevant micrometer (~0.13 mm) from the initial system. The characterisation of TL2 lateral misalignment (see Fig. S7) shows that although the lateral and axial distortions only change slightly from the initial system (see Table S3, mean distortion absolute difference from the initial system $< 0.3$ µm), laterally misaligning TL2 causes a decrease in the FOV over which a good image quality is achieved ($\geq 87\%$ reduction in the diffraction-limited volume, see Table S3). Hence, TL2 lateral alignment is an essential part in the remote-refocusing alignment procedure.

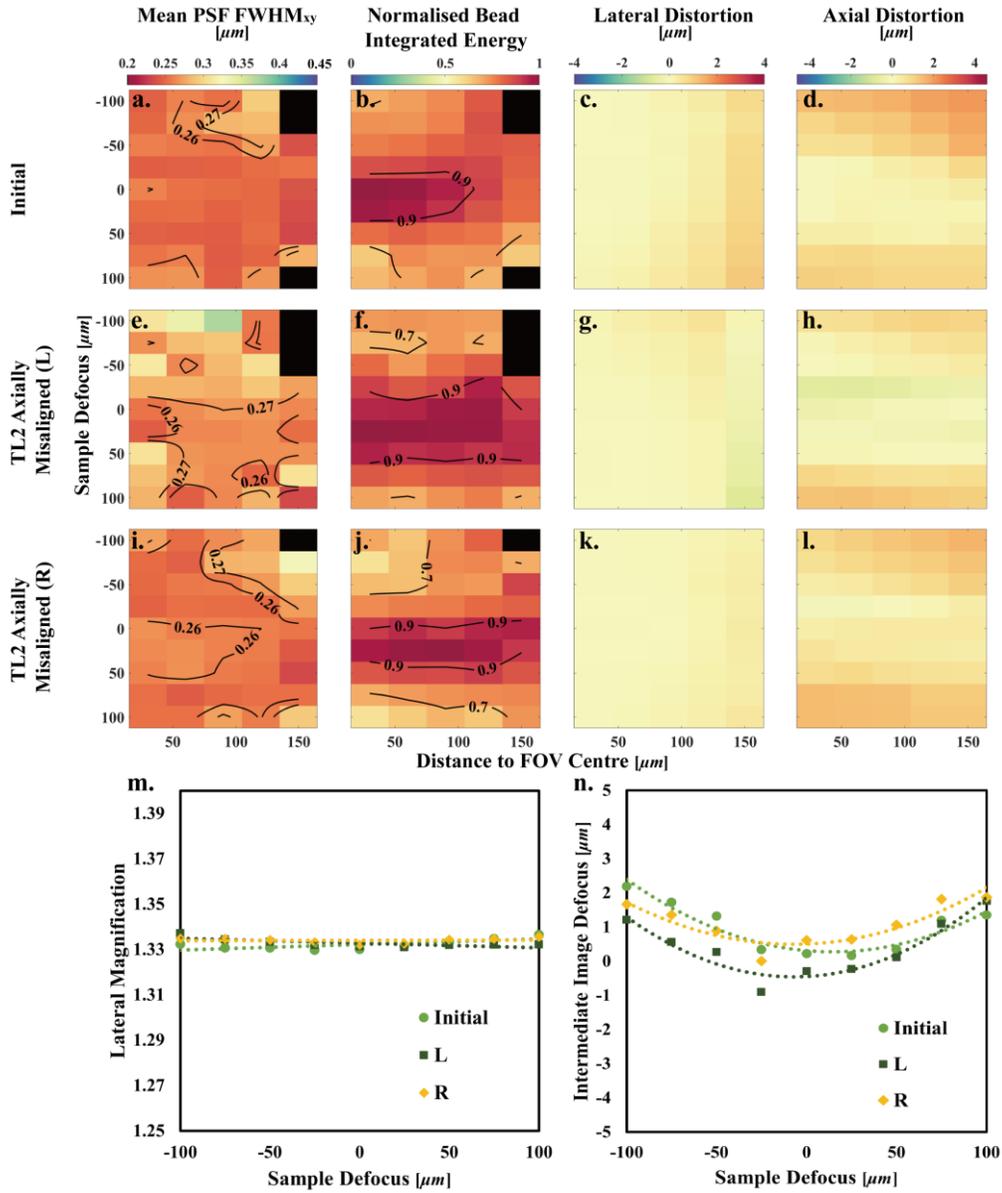

Fig. 3. Characterisation of TL2 axial misalignment. (a-d) Initial system. (e-h) TL2 axially misaligned by 2 mm towards TL1. (i-l) TL2 axially misaligned by 2 mm towards O2. (m) Lateral magnification as a function of sample defocus. The dashed lines are the linear fit lines to the measured data points. (n) Intermediate image defocus as a function of sample defocus. The dashed curves are the second-order polynomial fits to the measured data points.

### 3.3 Misalignment of TL2 focal length

In remote-refocusing systems, given the commercial tube lenses only provide specific focal lengths, TL2 can be formed by two achromatic doublets with an adjustable separation [12]. Different TL2 focal lengths were also tested to analyse the effect of overall magnification mismatch in remote-refocusing systems. The TL2 focal length was slightly increased (I) by increasing the separation of D1 and D2 (compared to initial alignment of TL2) by 2 turns of Thorlabs SM2 tube with pitch of 40 turns per inch, which is approximately 1.27 mm. The TL2 focal length was also decreased (D, 4 turns of SM2 tube from the initial TL2, which is approximately 2.54 mm), see Fig. 1. In both cases, the axial position of TL2 was then aligned relative to TL1 via the shear plate, and O2 was also adjusted to bring the variation in lateral magnification with defocus to less than $5\times10^{-5}$ μm$^{-1}$, as per the initial alignment procedure.

The measured lateral and axial magnifications of these two misaligned systems are in Table S1. When the focal length was increased compared to the initial system (I), the system lateral magnification decreased slightly from 1.3324 to 1.3323 as expected, although well within the measurement random error of 0.0005. In addition, the absolute difference between the lateral and axial magnifications (see Table S1) is larger than would be accepted by the alignment protocol (maximum 0.01). This misalignment leads to a slightly lower diffraction-limited region as depicted in the mean FWHM$_{xy}$ map and the diffraction-limited volume, decreased by 12% (see Fig. 4 (e) and Table S3). However, the volume over which the normalised bead integrated energy $\geq 0.8$ is slightly higher, increased by 32% (see Fig. 4 (f) and Table S4). Smaller lateral and axial distortions are also observed in Fig. 4 (g, h) and Table S3. Additionally, an increase in the system lateral magnification from 1.3324 to 1.3377 results in a larger decrease in the diffraction-limited region in the mean FWHM$_{xy}$ map (see Fig. 4 (i), decreased by 72% see Table S3). The normalised bead integrated energy $\geq 0.8$ also decreases by 1% (see Fig. 4 (j) and Table S4). Altering the focal length of TL2 also results in a modification of the system axial distortion, as depicted in Fig. 4 (h, l, n).

*3.4 Comparison with Zemax paraxial thin lens model*

The measurements presented above of the lateral magnification and intermediate image defocus as a function of sample defocus under axial misalignments of O1, O2 and TL2 were compared to simulations performed in Zemax OpticStudio, and the results are shown in Fig. S3.

In general, a reasonably good agreement is seen between experiment and simulation. To highlight the difference between experiment and simulation, Fig. S3 (m) shows the experimentally measured intermediate defocus after the simulated value has been subtracted. It can be seen that this residual distortion is approximately quadratic and is similar for all of the misalignments. Fig. S3 (n) shows the average and standard deviation of the data shown in panel (m). This quadratic distortion is predicted by the model of an ideal remote refocusing system presented Botcherby et al. [19], see the expression for δz in equation 21.

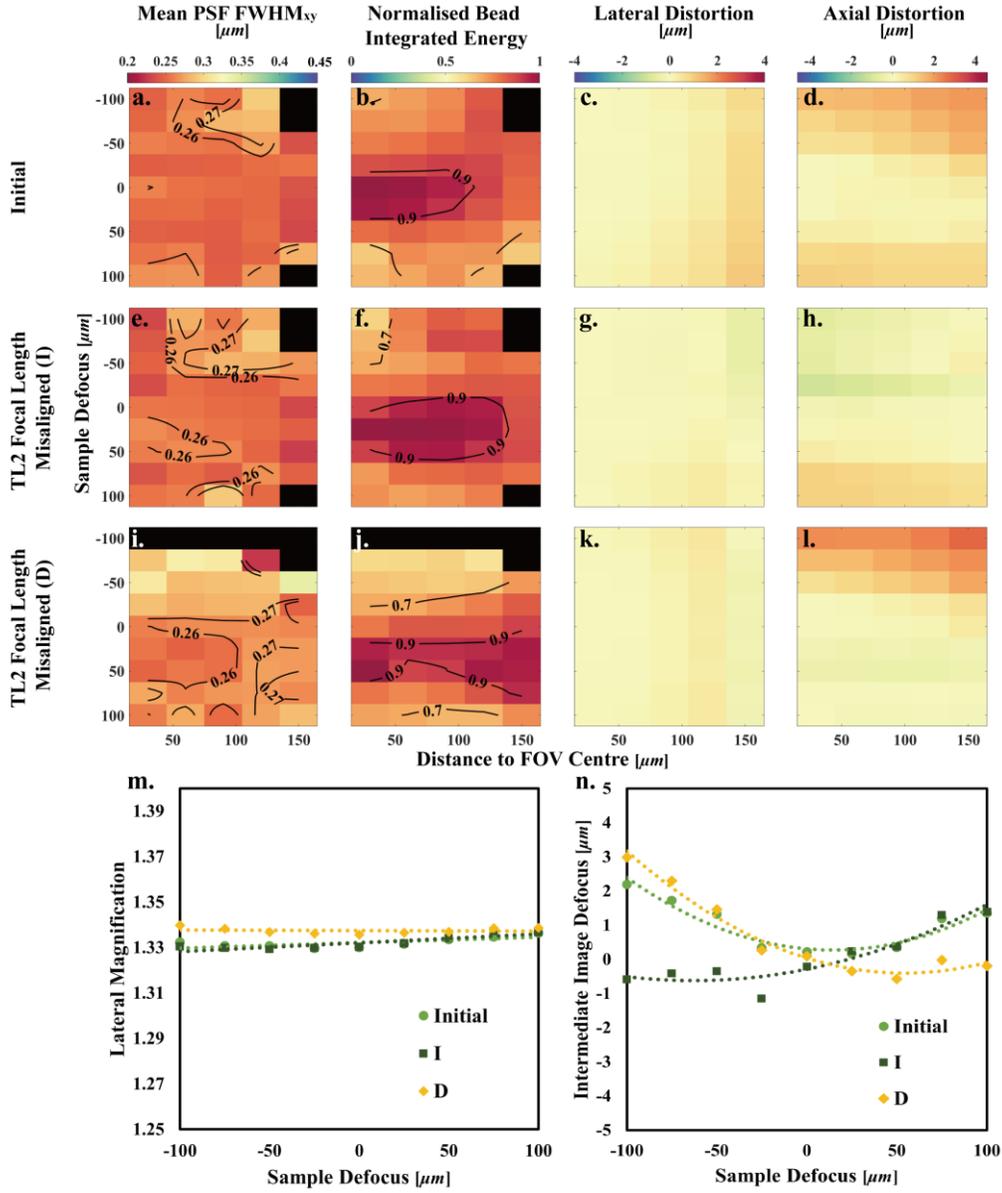

Fig. 4. Characterisation of TL2 focal length misalignment. (a-d) Initial system. (e-h) TL2 with increased focal length. (i-l) TL2 with decreased focal length. (m) Lateral magnification as a function of sample defocus. The dashed lines are the linear fitting lines to the measured data points. (n) Intermediate image defocus as a function of sample defocus. The dashed curves are the second-order polynomial fits to the measured data points.

## 4. Discussion

### 4.1 Lateral magnification

The ideal lateral magnification of the refocusing system should theoretically be equal to the ratio of refractive indices between the sample and remote spaces. For example, Mohanan and Corbett [9] showed that a 1% magnification mismatch leads to a 50% reduction in the remote-refocusing range. For the system here, the lateral magnification should be equal to the refractive index of water. In the literature, the refractive index of water is reported for 20°C and 1 bar as

1.33626 at 515 nm, and 1.33348 at 589 nm [20]. We measured the refractive index of the immersion water used to be 1.3335±0.0001 for a wavelength of 589 nm, which is consistent with the literature value. All experiments performed here were carried out using wavelengths over the range 500-550 nm. Linear interpolation of the values stated above from reference [19] gives a refractive index for water of 1.3359 at the centre wavelength used here of 525 nm.

The lateral magnification of our initial alignment was 1.3324 with an estimated random error of ±0.0005 and an estimate of the maximum systematic error of ±0.0005. The lateral magnification of the initial alignment differs from the literature value of the refractive index of water (1.3359) by more than the overall experimental error. This is potentially due to the interplay of aberrations between the 4 lenses in the experimental system (O1, TL1, TL2, O2), our choice of alignment procedure aiming to match the lateral and axial magnifications rather than aiming to just reach a specific lateral magnification, or a systematic error in our measurement of the lateral magnification that we did not account for. The condition TL2 (D) has a lateral magnification (1.3377) closer to the literature value of the refractive index, but this alignment had a reduced diffraction-limited volume (72%) compared to the initial condition, which would suggest that the problem is not our choice of alignment procedure. Further work is required to understand this.

As pointed out by theoretical calculations by Mohanan and Corbett, see figure 6 of reference [9], small variations in the lateral magnification on the order of 0.995-1.005 can lead to defocus as a function of refocus distance (axial distortion) that partially compensates for spherical aberrations and increases the diffraction-limited refocus range when considering points along optical axis. These variations in lateral magnification are expected to lead to a shift in where the best imaging performance is obtained to one side of the focal plane of O1. However, in our initial alignment the variation of axial distortion as a function of sample defocus is reasonably symmetrical about zero sample defocus, see Fig. 4 (d & n), suggesting a good mapping of the pupil of O1 to the pupil of O2. When increasing the focal length of TL2 (I), we see the position of minimum axial distortion shift to the left in Fig. 4(n), and when decreasing the focal length of TL2 (D) we see a shift to the right. This is broadly consistent with what is expected based on figure 6 of [9] as small amounts of defocus (distortion) compensate for spherical aberration.

In the future, an accurately produced fluorescent 3D test sample consisting of a known 3D distribution of point sources with the desired refractive index – such as that presented by Corbett et al. [21] – would enable the distortion and aberration in the system to be quantified by only requiring a scan of Stage 2, with Stage 1 remaining stationary.

*4.2 Collection efficiency as a function of remote-refocus distance*

The collection efficiency of remote-refocusing systems as a function of remote-refocus distance on-axis has previously been reported by Kim et al. [14], see Supplementary Figure 4 (c). We measured the volume over which the normalised bead integrated energy was $\geq 0.8$. To put these results in context, it is useful to consider geometrically the volume over which the optical system is expected to be able to collect all of the fluorescence signal. In the Supplementary Material section 2, we build on the approach of Kumar et al. [22] which gives a predicted value of $8.0 \times 10^6$ μm$^3$ for the system used here and has a good order of magnitude agreement with the values reported in Table S4 (initial alignment $7.1 \times 10^6$ μm$^3$).

## 5. Conclusion

This paper presents an analysis of the effects of misalignment on a remote-refocusing system employing a 60×/1.2NA water Nikon O1 lens and a 50×/0.95NA air Olympus O2 lens. TL1 was the ITL200 from Thorlabs and TL2 was formed from a pair of achromatic doublets (322305000, Linos and AC508-300-A, Thorlabs). We performed an initial alignment of the system and then studied the effect of axial misalignments of O1 and O2, axial misalignment of TL1 relative to TL2, lateral misalignments of TL2 and axial misalignment of the separation of the two doublets forming TL2 causing a change in the focal length of TL2. For each instance of the setup, we measured the FWHM and the normalised integrated energy of 100 nm fluorescent beads and the axial and lateral distortion of the system: all of these quantities were mapped over the remote-refocusing range and as a function of lateral image position.

Axial movement of O1 by ±1 mm or O2 by ±2 mm caused a reduction in the diffraction-limited imaging volume in the range 65 to 73%. Misalignment of O1 towards TL1 direction resulted in similar effects to misalignment of O2 towards TL2. Axial misalignments of TL1 relative to TL2 by ±2 mm reduced the diffraction-limited imaging volume by (75% (L) and 19% (R)). The misalignment of TL1 relative to TL2 is slightly larger than the measured alignment error using a collimated laser diode and shear plate of 1.54 mm, and therefore we believe that alignment using this method is a reasonable approach. The measurements of the lateral magnification and intermediate image defocus as a function of sample defocus under axial misalignments of O1, O2 and TL2 were found to be well described by paraxial thin lens simulations performed in Zemax apart from a residual weak quadratic dependence of intermediate image defocus as a function of sample defocus that is predicted by the model of the ideal system by Botcherby et al. [19]. These measurements also help indicate which elements require axial adjustment. Lateral motion of TL2 by ~0.13 mm caused the biggest reduction diffraction-limited imaging volume in the range 87 to 93%.

The initial alignment was found to give the overall best performance of the system in terms of the diffraction-limited volume ($9.6 \times 10^6$ μm$^3$). Some of the other conditions had an increased volume where the normalised bead integrated energy was ≥ 0.8. But overall, our results were consistent with our alignment protocol giving the best performance.

This paper provides a characterisation of the errors due to misalignment for a particular remote-refocusing system. Further work is required to check if these findings are generally seen in systems employing different combinations of optical elements. Further work is also required to determine whether the alignment procedure presented here provides the best performance for other combinations of lenses in a remote-refocus system.

**Funding**

The authors gratefully acknowledge funding from UK Engineering and Physical Sciences Research Council grants EP/T003103/1 and EP/R511547/1, and a CRUK Accelerator award (A29368).

**Acknowledgements**

The authors gratefully acknowledge the expert help from Martin Kehoe and Simon Johnson in the Optics Workshop of the Photonics Group of Imperial College London who helped fabricate the custom components for this remote-refocusing test rig system; Terry Wright for providing his STM pinhole finding and lateral magnification algorithm for our adaptation; Liuba Dvinskikh for valuable advice and discussions on the bead psf fitting procedure; Edwin Garcia Castano for valuable assistance with the bead sample preparation.

**Disclosures**

CD has filed a patent application on dual-view oblique plane microscopy and has a licensed granted patent on oblique plane microscopy, which both utilise remote-refocusing systems.

**Data availability**

Data underlying the results presented in this paper are available in Data File 1.

**Supplemental document**

See Supplementary Material for supporting content.

# Alignment and Characterisation of Remote-Refocusing Systems: Supplementary Material

## 1. Sample Preparation

A precision #1.5 glass coverslip (630-2186, Marienfield) was cleaned with acetone for 5 minutes, ethanol for 5 minutes and washed three times with water. Then 2 µL 1:100 dilution of poly-L-lysine (P8920, Sigma-Aldrich) in water was incubated on the coverslip at room temperature for 5 minutes. Excess poly-L-lysine was removed by three washes with water. 100 nm fluorescence beads (T7279, TetraSpeck) were diluted 500-fold in water and a drop (50 µL) placed on the poly-L-lysine-coated coverslip for 10 minutes. Unbound beads were removed by washing three times in water. The beads were then covered with polyvinyl acetate (MOWIOL® 4-88 Reagent, 475904 Millipore, Mowiol) and a Superfrost Plus Adhesion microscope slide (J18000AMNZ, Epredia) was placed on top. The coverslip and microscope slide were sealed permanently with silicone sealant (Polycraft ZA22, Mould RTV Addition Cure Mould Making Silicone Rubber, MBFibreglass). The slide was incubated overnight at 37°C to cure the mounting media and silicone.

## 2. Supplementary Text

### 2.1 Estimation of Theoretical Diffraction-Limited Lateral Resolution

The estimated ideal point spread function (PSF) is obtained using the convolution of the scalar Airy distribution, a spherical 100 nm diameter bead where the fluorophore is assumed to be uniformly distributed through the sphere, and a square pixel. This can be written as

$$\text{PSF}_{\text{airy\_bead}}(x,y) = \text{Airy}_{\text{ideal}}(x,y) * \text{Bead}_{\text{ideal}}(x,y) * \text{Pixel}(x,y), \quad (S1)$$

$$\text{Airy}_{\text{ideal}}(x,y) = \{2J_1[(2\pi/\lambda)\text{NA}R]/[(2\pi/\lambda)\text{NA}R]\}^2, \quad (S2)$$

$$\text{Bead}_{\text{ideal}}(R) = 2(R_{\text{bead}}^2 - R^2)^{1/2} \text{ for } R \leq R_{\text{bead}}$$
$$= 0 \text{ for } R > R_{\text{bead}} \quad (S3)$$

$$\text{Pixel}(x,y) = 1 \text{ for } |x| \leq P/2 \text{ and } |y| \leq P/2$$
$$R = (x^2+y^2)^{1/2}, \quad (S4)$$

where $R_{\text{bead}}$ is the 50 nm bead radius and $P$ is the width of the pixel. This provides an FWHM of 0.23 µm.

The Strehl criterion is used commonly in optics to provide a condition where an optical system can be considered to be practically diffraction limited [1]. A system is considered to be diffraction limited by this criterion if the peak of the actual PSF is greater than or equal to 80% of the PSF that would be obtained if the system was aberration free. To make use of this criterion in the context of the PSF FWHM measured in this work, we considered how much a Gaussian PSF would need to be broadened by (whilst keeping the integrated PSF energy constant) for the peak intensity to drop to 80% of the original value. This results in a broadening factor of $0.8^{-1/2}$. Therefore, the broadened bead FWHM that still meets the Strehl condition is 0.26 µm and we chose to use this to define the region over which the system can be considered to be practically diffraction limited.

### 2.2 Geometrical Remote-Refocusing Volume

We consider geometrically the volume where rays pass within both the limiting field of view of the system $D_{\text{lim}}$ and the limiting angular aperture of the system $\theta_{\text{lim}}$, see Fig. S8.

$D_{lim}$ is given by the minimum of the field number of O1, $FN_{O1}$, and the field number of O2, $FN_{O2}$, projected into sample space, i.e., $D_{lim} = \min\{FN_{O1}/M_{O1}, FN_{O2}/M_{O2}/M_{total}\}$. For the system used here, $D_{lim} = 398$ μm, $M_{total} = 1.333$, and $M_{O1}$ and $M_{O2}$ are the magnifications of O1 and O2.

$\theta_{lim}$ is given by the minimum angular acceptance of O1 and O2, i.e., $\theta_{lim} = \min\{\arcsin(NA_{O1}/n_1), \arcsin(NA_{O2}/n_2)\}$. For the system used here $\theta_{lim} = 64°$.

The maximum geometric refocus distance, $Z_{max}$, occurs when rays emitted at the limiting angle $\theta_{lim}$ from a point object on the optical axis away from the objective focal plane just pass through the limiting field of view $D_{lim}$ in the focal plane, see Fig. S8. Using trigonometry, $Z_{max}$ is given by the equation

$$Z_{max} = D_{lim}/(2 \tan \theta_{lim}). \tag{S6}$$

For the system used here, $Z_{max} = 95$ μm.

The geometric region over which all rays emitted within the collection cone of the remote refocussing system defined by $\theta_{lim}$ is therefore described by a double-ended cone with diameter $D_{lim}$, tip-to-tip distance $2Z_{max}$ and volume

$$V_{geom} = \pi D_{lim}^2 Z_{max}/6. \tag{S7}$$

For the system used here, $V_{geom} = 8.0 \times 10^6$ μm³. This is in good order-of-magnitude agreement with the values reported below in supplementary Table S4.

*2.3 Simulations of O1, O2 and TL2 axially misalignments*

## 3. Supplemental Figures

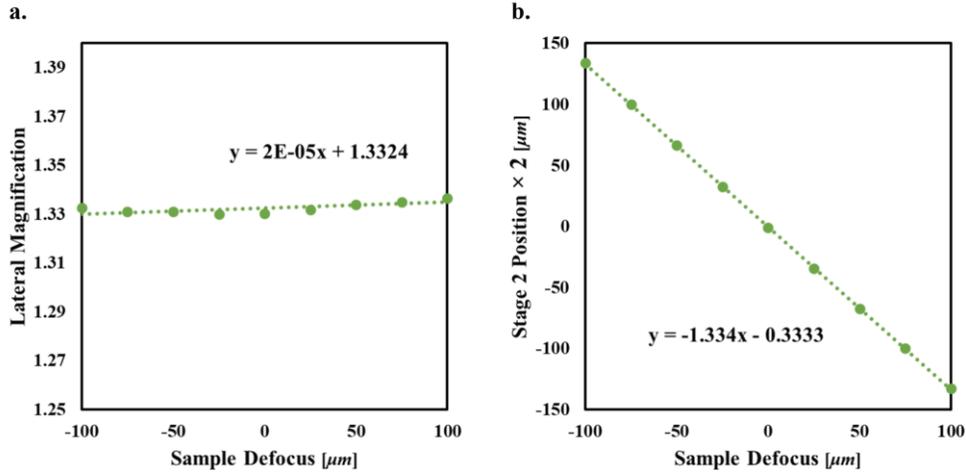

Fig. S1. (a) Linear fit to the measured lateral magnifications within the whole refocusing range. (b) Linear fit to twice the Stage 2 position for the in-focus image as a function of sample defocus. The axial magnification equals to (-) gradient. At each sample defocus, the lateral and axial magnifications were measured from the Stage 2 position closest to the sample being in focus on C2.

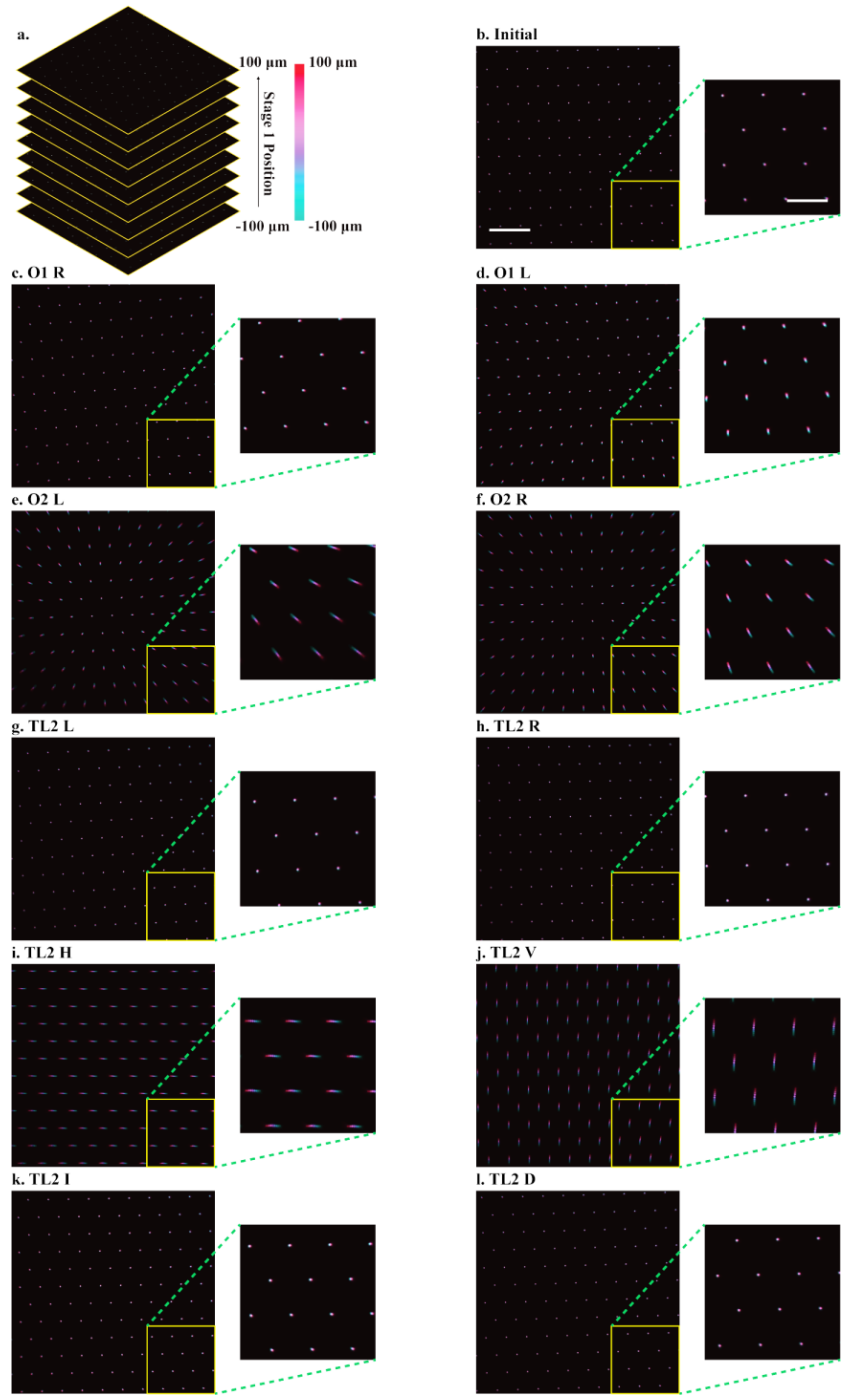

Fig. S2. Color-coded maximum intensity projection (MIP) of 9 in-focus STM images in all tested cases. (a) Cartoon illustrating the 9 in-focus STM images and corresponding axial range and false-colour bar scale. (b-l) Color-coded MIPs across the whole refocusing range for the initial condition (b) and different misalignments (c-l) of the full FOV (2304 pixels × 2304 pixels; scale bar, 40 μm); zoomed-in images on the right of each panel (768 pixels × 768 pixels; scale bar, 20 μm) show the region indicated by the yellow squares. O, objective; TL, tube lens; L, left; R, right; H, horizontal; V, vertical; I, increase; D, decrease. See Fig. 1 for diagram showing directions of motion for each misalignment.

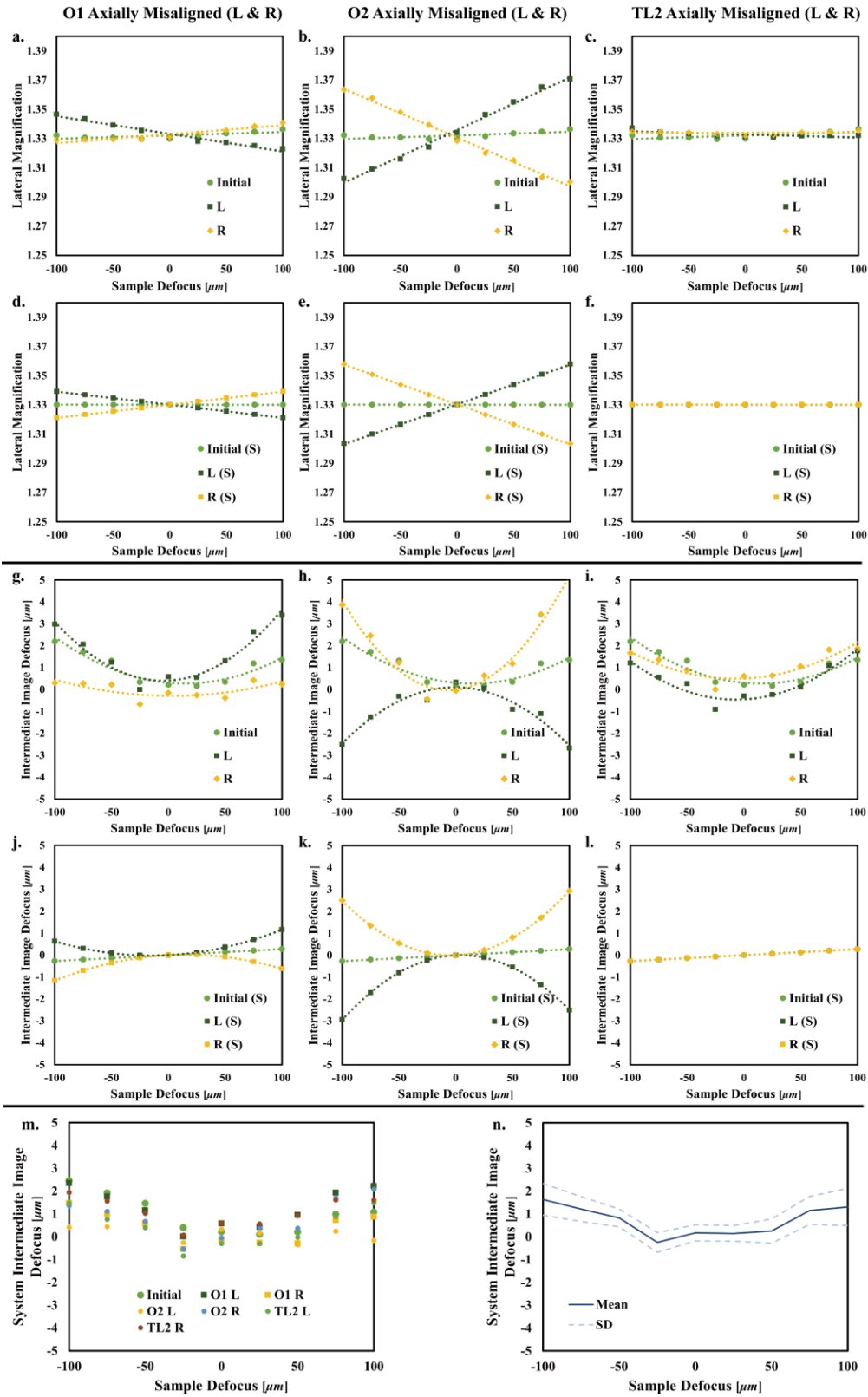

Fig. S3. Comparisons of experimental measurements of lateral magnification and intermediate image defocus with the simulations performed in Zemax using paraxial thin lenses. (a-c, g-i) Experimental measurements of lateral magnification and intermediate image defocus as functions of sample defocus. (d-f, j-l) Simulation results of lateral magnification and intermediate image defocus as functions of sample defocus. (a-c) are the same as figures shown in Fig. S5 (m), Fig. 2 (m) and Fig. 3 (m). (g-i) are the same figures in Fig. S5 (n), Fig. 2 (n) and Fig. 3 (n). (m) Data points obtained by subtracting the simulation results from the corresponding experimental measurements of intermediate image defocus. (n) Average of the values obtained from the different misalignments shown in (m). The two dashed lines indicate plus and minus one standard deviation of the different misalignments in (m). L, left; R, right; S, simulation; SD, standard deviation.

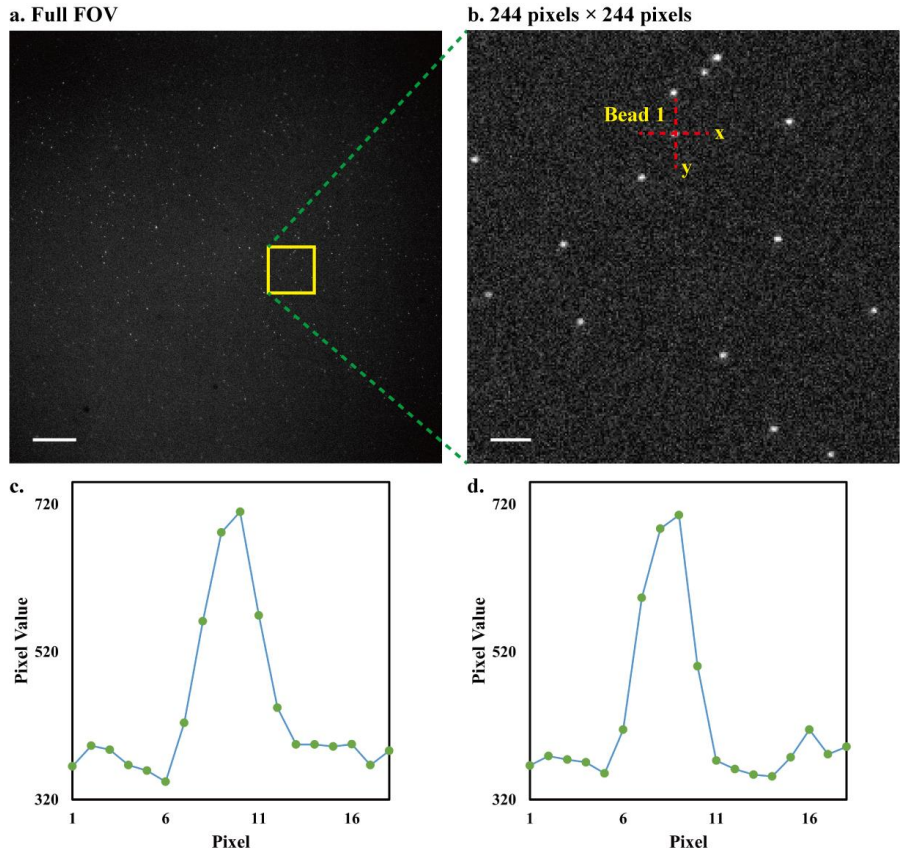

Fig. S4. (a) An example image of 100 nm beads acquired with the initial system at 0 μm defocus position. Scale bar, 20 μm. (b) A zoom-in image of the yellow square in (a). Scale bar, 2 μm. (c) and (d) are respectively horizontal and vertical profiles of Bead 1 indicated with red dashed lines in (b). Gaussian fits to the profiles of beads are shown in blue dashed curves. The FWHM values of curves shown in (c) and (d) are 0.292 μm and 0.273 μm respectively (mean FWHM = 0.283 μm).

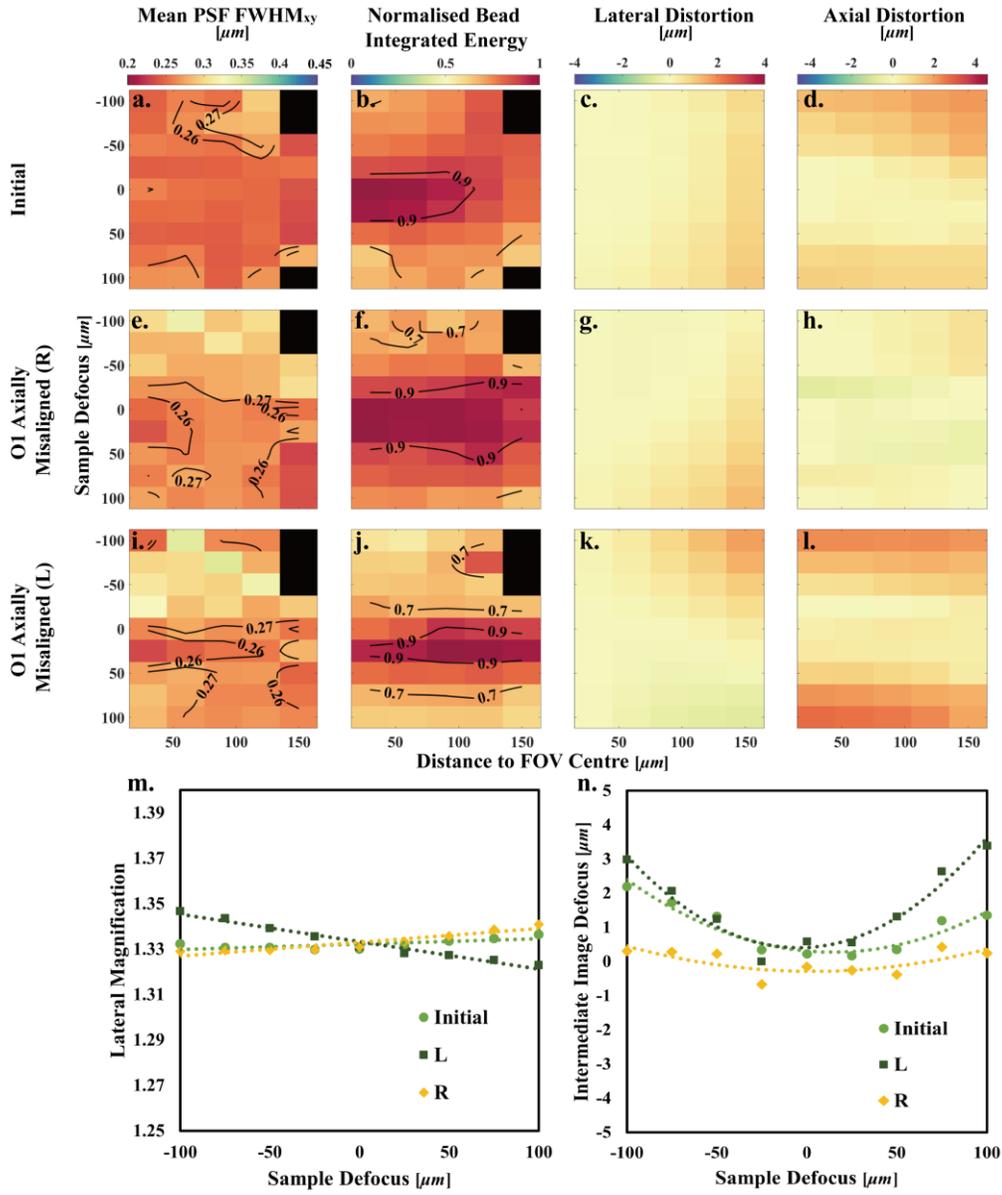

Fig. S5. Characterisation of O1 axial misalignment. (a-d) Initial system. (e-h) O1 axially misaligned by 1 mm towards TL1. (i-l) O1 axially misaligned by 1 mm towards Stage 1. (m) Lateral magnification as a function of sample defocus. The dashed lines are linear fits to the measured data points. (n) Intermediate image defocus as a function of sample defocus. The dashed curves are the second-order polynomial fits to the measured data points.

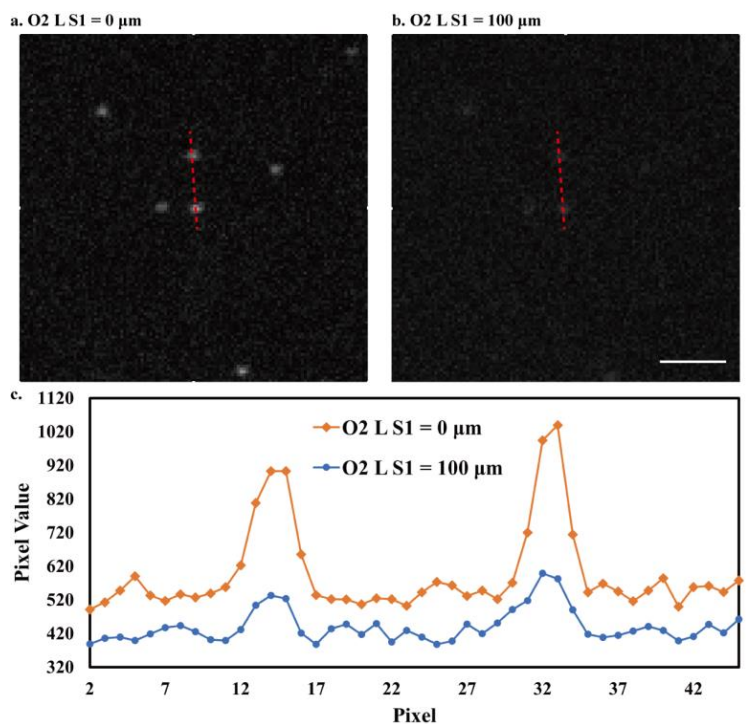

Fig. S6. (a) Image of two 100 nm beads (average 40 μm from the centre of FOV) acquired for misalignment of O2 for a Stage 1 position of 0 μm. (b) Same field of view and conditions except now for a Stage 1 position of 100 μm. In both cases the optical system has been refocused using M2. Scale bar, 2 μm. (c) Line profiles along the red dashed lines indicated in (a) (orange) and (b) (blue). O, objective; L, left; S1, Stage 1.

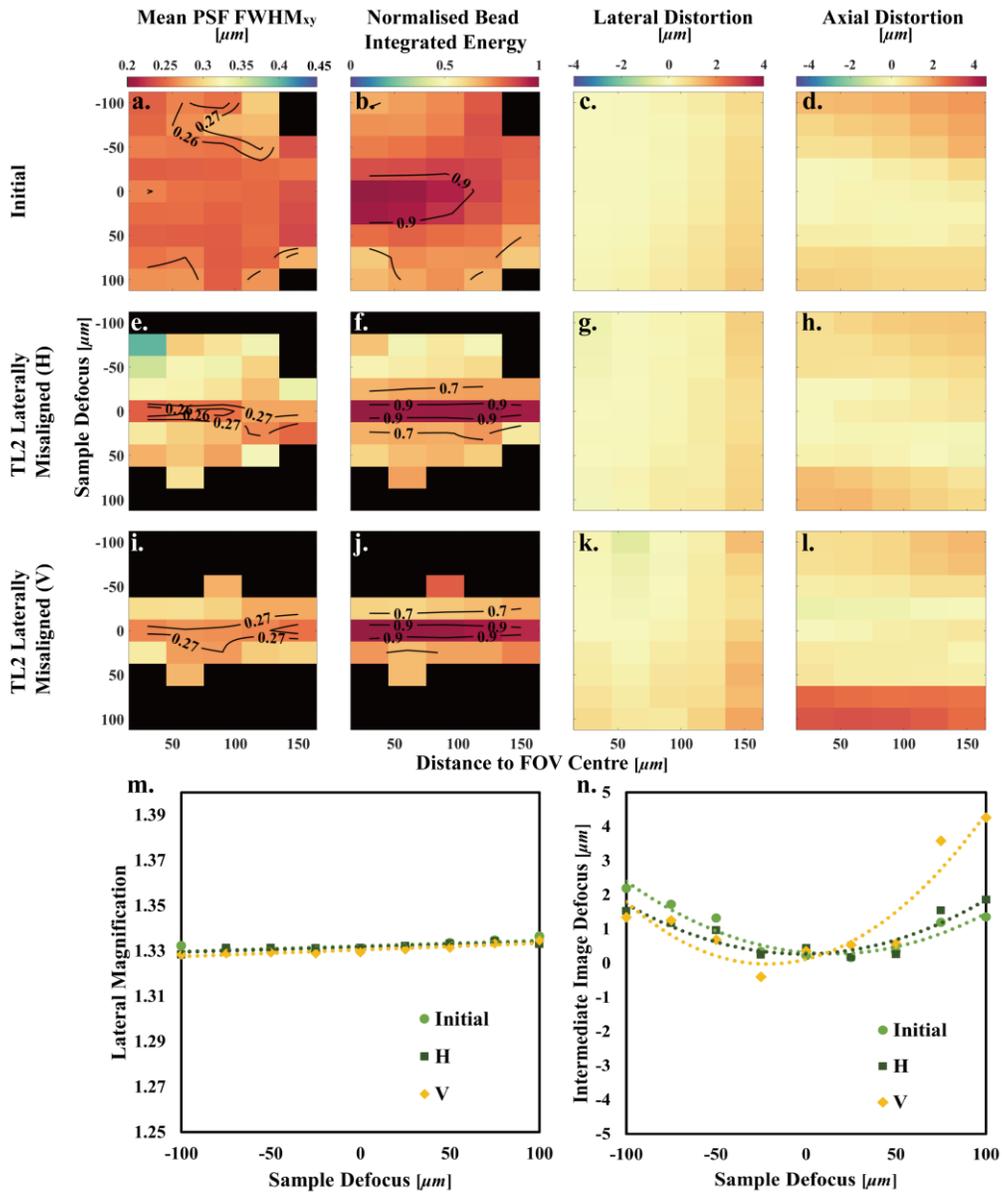

Fig. S7. Characterisation of TL2 lateral misalignment. (a-d) Initial system. (e-h) TL2 laterally misaligned in horizontal direction. (i-l) TL2 axially misaligned in vertical direction. (m) Lateral magnification as a function of sample defocus. The dashed lines are the linear fitting lines to the measured data points. (n) Intermediate image defocus as a function of sample defocus. The dashed curves are the second-order polynomial fits to the measured data points.

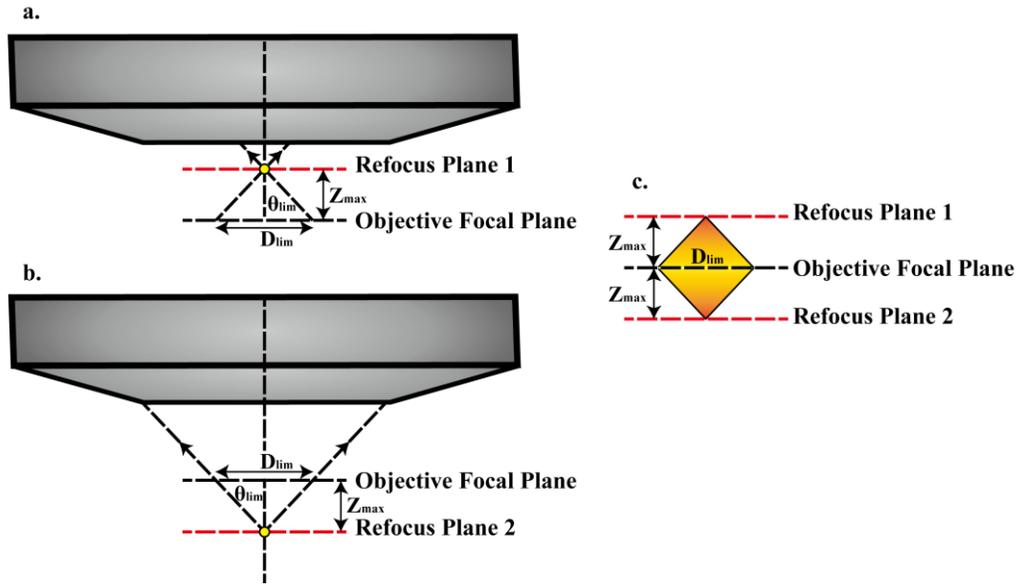

Fig. S8. Schematic showing the geometric maximum refocusing distance from the objective focal plane $Z_{max}$. (a) shows the closest refocus plane. (b) shows the furthest refocus plane. $D_{lim}$, limiting FOV of refocus system; $\theta_{lim}$, limiting angular acceptance of the system; $Z_{max}$, distance at which geometric FOV shrinks to zero. (c) shows the side projection of the geometric maximum refocusing volume.

## 4. Supplemental Tables

Table S1. Lateral and axial magnifications in all tested cases.

| Case | Lateral Magnification (FOV centre) | Axial Magnification | Abs (Lateral Magnification - Axial Magnification) * |
|---|---|---|---|
| Estimated random error | ±0.0005 | ±0.0050 | |
| Maximum estimated systematic error | ±0.0005 | | |
| Initial | 1.3324 | 1.334 | 0.0016 |
| O1 (R) | 1.3332 | 1.329 | 0.0039 |
| O1 (L) | 1.3334 | 1.325 | 0.0087 |
| O2 (L) | 1.3360 | 1.329 | 0.0067 |
| O2 (R) | 1.3308 | 1.324 | 0.0068 |
| TL2 (L) | 1.3330 | 1.325 | 0.0083 |
| TL2 (R) | 1.3342 | 1.325 | 0.0095 |
| TL2 (H) | 1.3320 | 1.327 | 0.0053 |
| TL2 (V) | 1.3307 | 1.315 | 0.0160 |
| TL2 (I) | 1.3323 | 1.315 | 0.0176 |
| TL2 (D) | 1.3377 | 1.343 | 0.0056 |

*: Absolute value.

**Table S2. Gradients of measured lateral magnification lines in all tested cases.**

| Case | Gradient of Lateral Magnification Line ($\mu m^{-1}$) $\pm 8\times10^{-6}$ |
|---|---|
| Initial | $2\times10^{-5}$ |
| O1 (R) | $6\times10^{-5}$ |
| O1 (L) | $-1\times10^{-4}$ |
| O2 (L) | $4\times10^{-4}$ |
| O2 (R) | $-3\times10^{-4}$ |
| TL2 (L) | $-2\times10^{-5}$ |
| TL2 (R) | $1\times10^{-6}$ |
| TL2 (H) | $2\times10^{-5}$ |
| TL2 (V) | $3\times10^{-5}$ |
| TL2 (I) | $4\times10^{-5}$ |
| TL2 (D) | $-2\times10^{-6}$ |

**Table S3. Mean absolute lateral and axial distortions and diffraction-limited volume in all tested cases.**

| Case | Mean absolute lateral distortion* ($\mu m$) | Mean absolute axial distortion** ($\mu m$) | Diffraction-limited volume ($\times 10^6\ \mu m^3$) | Percentage difference in diffraction-limited volume from Initial*** (%) |
|---|---|---|---|---|
| Initial | 0.31 | 0.80 | 9.6 | 0 |
| O1 (R) | 0.28 | 0.24 | 3.4 | -65 |
| O1 (L) | 0.38 | 1.21 | 3.1 | -68 |
| O2 (L) | 1.04 | 0.77 | 2.7 | -73 |
| O2 (R) | 1.01 | 1.49 | 2.7 | -72 |
| TL2 (L) | 0.17 | 0.54 | 2.4 | -75 |
| TL2 (R) | 0.13 | 0.85 | 7.8 | -19 |
| TL2 (H) | 0.30 | 0.68 | 1.3 | -87 |
| TL2 (V) | 0.49 | 1.09 | 0.6 | -93 |
| TL2 (I) | 0.13 | 0.47 | 8.4 | -12 |
| TL2 (D) | 0.17 | 0.74 | 2.7 | -72 |

*: Average over 2D histograms shown in panels c, g, k of figures 2, 3, 4, S5 and S7.
**: Average over 2D histograms shown in panels d, h, l of figures 2, 3, 4, S5 and S7.
***: Percentage Difference from Initial = (Diffraction-limited volume - Diffraction-limited volume of Initial)/(Diffraction-limited volume of Initial)×100.

**Table S4. Volume where normalised bead integrated energy is ≥ 0.8 in all tested cases.**

| Case | Volume where normalised bead integrated energy is ≥ 0.8 (×10$^6$ μm$^3$) | Percentage difference in volume from Initial* (%) |
|---|---|---|
| Initial | 7.1 | 0 |
| O1 (R) | 9.3 | 31 |
| O1 (L) | 5.7 | -20 |
| O2 (L) | 6.8 | -4 |
| O2 (R) | 4.5 | -36 |
| TL2 (L) | 9.4 | 32 |
| TL2 (R) | 7.3 | 3 |
| TL2 (H) | 1.8 | -75 |
| TL2 (V) | 2.1 | -70 |
| TL2 (I) | 9.4 | 32 |
| TL2 (D) | 7.0 | -1 |

*: Percentage Difference from Initial = (Volume without energy loss - Initial volume without energy loss)/(Initial volume without energy loss)×100.